%% file: MBCB-crowdSRV2.tex
\documentclass[aps,prl,twocolumn,showpacs,superscriptaddress,groupedaddress]{revtex4-1}  % for review and submission
\usepackage{graphicx}
\usepackage{anysize}
\usepackage{lineno}
\usepackage[amssymb]{SIunits}
\usepackage{lipsum}
\usepackage{amsmath}
\usepackage{amssymb}
\usepackage{SIunits}
\let\revappendix\appendix
\usepackage[capitalize]{cleveref}
\newcommand \be {\begin{equation}}
\newcommand \ee {\end{equation}}
\newcommand \bea {\begin{eqnarray}}
\newcommand \eea {\end{eqnarray}}

\begin{document}

%\preprint{}

\title{Extreme congestion of microswimmers at a bottleneck constriction}

% The following information is for internal review, please remove them for submission
%\widetext
%\leftline{Version xx as of \today}
%\leftline{Primary authors: Joe E. Physics}
%\leftline{To be submitted to (PRL, PRD-RC, PRD, PLB; choose one.)}
%\leftline{Comment to {\tt d0-run2eb-nnn@fnal.gov} by xxx, yyy}
%\centerline{\em D\O\ INTERNAL DOCUMENT -- NOT FOR PUBLIC DISTRIBUTION}

% the following line is for submission, including submission to the arXiv!!
%\hspace{5.2in} \mbox{Fermilab-Pub-04/xxx-E
\input author_list.tex 


\date{\today}

%\pacs{}
%\keywords{}
\begin{abstract}

When attracted by a stimulus (e. g. light), microswimmers can build up very densely at a constriction and thus cause clogging. The micro-alga \textit{Chlamydomonas Reinhardtii} is used here as a model system to study this phenomenon. Its negative phototaxis makes the algae swim away from a light source and go through a microfabricated bottleneck-shaped constriction. Successive clogging events interspersed with bursts of algae are observed. A power law decrease is found to describe well the distribution of time lapses of blockages. Moreover, the evacuation time is found to increase when increasing the swimming velocity. These results might be related to the phenomenology of crowd dynamics and in particular what has been called the Faster is Slower effect in the dedicated literature. It also raises the question of the presence of tangential solid friction between motile cells densely packed that may accompany arches formation. Using the framework of crowd dynamics we analyze the microswimmers behavior and in particular question the role of hydrodynamics.
%which is poorly documented in literature.

\end{abstract}
\maketitle

%\linenumbers
%\modulolinenumbers[10] %Cada cuanto el conteo de las l�?neas

In bio-technologies, harvesting micro-organisms such as bacteria \cite{matsumura2018microbial} or micro-algae \cite{milledge2013review} is now a tremendous challenge to extract new drugs and chemicals \cite{nakagawa2016total,deloache2015enzyme}. Several microfluidic devices have recently emerged to achieve such a task \cite{honsvall2016continuous,karimi2015interplay,cabrera2001continuous}.  A clogging phase can often occur when concentrated cells are transported by a fluid \cite{dressaire2017clogging, hassanpourfard2016dynamics, wyss2006mechanism}.
At the microscale, clogging results from the accumulation of cells at a constriction, leading to the build-up of fouling layers. Clogging can severely disrupt the performance of microfluidic devices \cite{mukhopadhyay2005microfluidic}. Initially, it reduces the permeability of channels or pores and eventually lead to a complete obstruction of the flow \cite{griffiths2014combined, marty2012formation}. Clogging is thus one of the leading causes of efficiency loss in high-throughput technologies processes \cite{wolff2003integrating}. To remedy fouling and clogging, it is usually necessary to completely stop the process and use cleaning strategies that consume energy and time before restarting the process \cite{lim2003membrane}. Motile cells (bacteria or microalgae) can self-accumulate without an external flow and can go to nest in a particular place of the circuit (pores, porous regions, constrictions) thus forming an active clogging that is essential to characterize \cite{marty2012formation}. Beyond clogging, this type of accumulation can be a precursor to the formation of a biofilm. 
\begin{figure*}[!ht]
\centering
  \includegraphics[width=2\columnwidth]{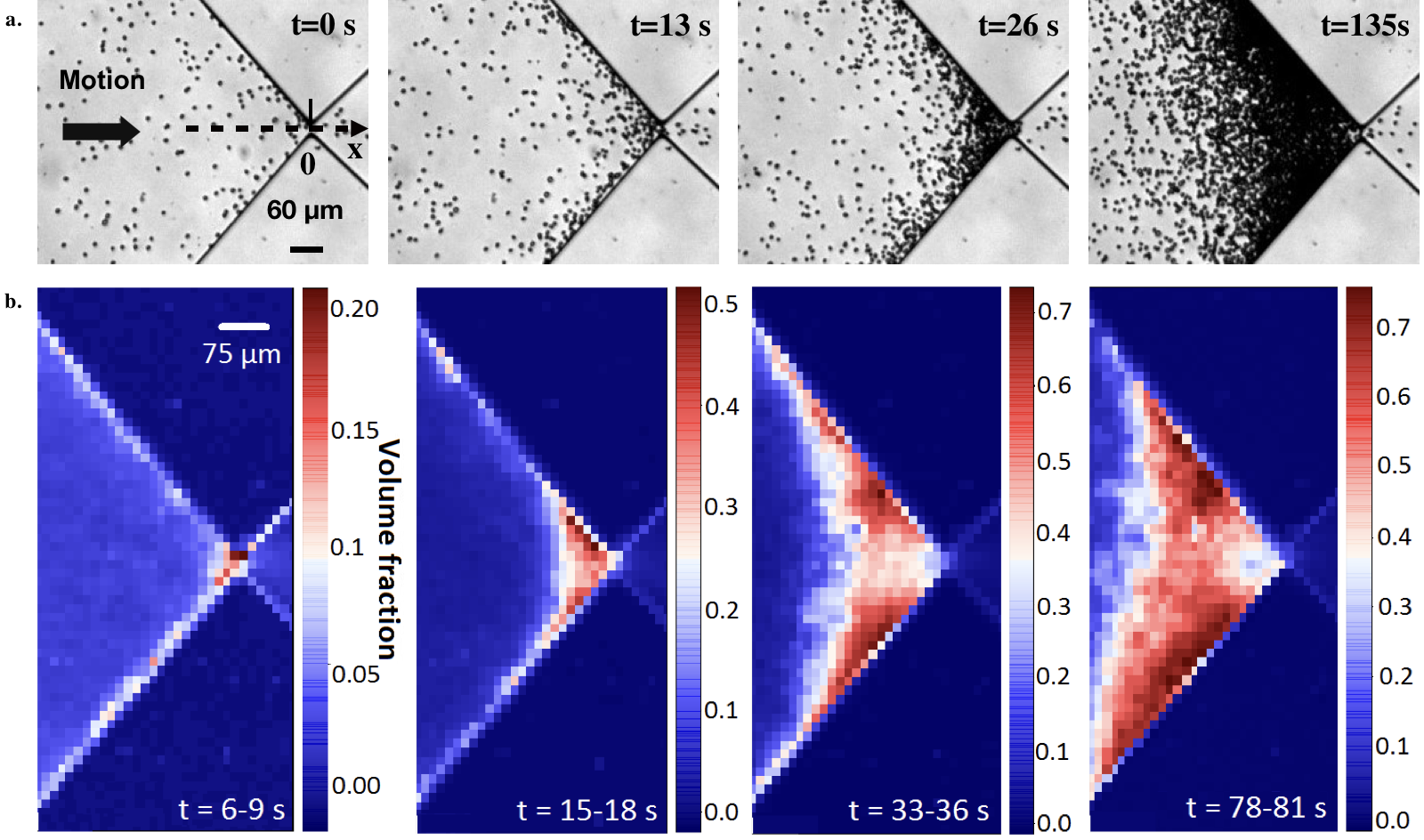}  	   
  \caption{a. Snapshots of clogging process through a $\unit{15}{\micro\meter}$ constriction at different times. The microswimmers flee the light source set on the left side of the bottleneck. b. Colormaps of the volume fraction in microswimmers averaged over 3 seconds at different stages of clogging.}  \label{snapshots}
\end{figure*}

More generally, clogging problems occur in extremely varied situations that are nowadays very much studied in the context of crowd dynamics. Crowd motion modeling has become a very active field of research, and deep debates are still ongoing in the community on how to model certain behaviors such as panic movements \cite{helbing2000simulating}. In this particular case, the Faster is Slower (FiS) effect plays a central role: it essentially indicates that, in certain situations where the crowd is very dense, the willingness of individuals is to increase their speed in their effort to evacuate, which has the effect of blocking the entire evacuation process.  In addition, it has been established that these paradoxical effects observed in crowd dynamics persist in a very wide variety of systems: in a granular silo, in a suspension as well as in a flock of sheep or in a crowd of pedestrians \cite{zuriguel2014clogging}. A physicist's approach therefore invites us to look for a common physical mechanism behind these observable measured values. At a minimum, it would be desirable to succeed in defining universality classes in this family of problems. In particular, a class of system could be that of active suspensions which intrinsically involves hydrodynamic interactions between agents. How microhydrodynamics will affect crowd dynamics is the question that we like to raise in this work.

\begin{figure}[!ht]
  %\centering
  \begin{center}
\includegraphics[width=\columnwidth]{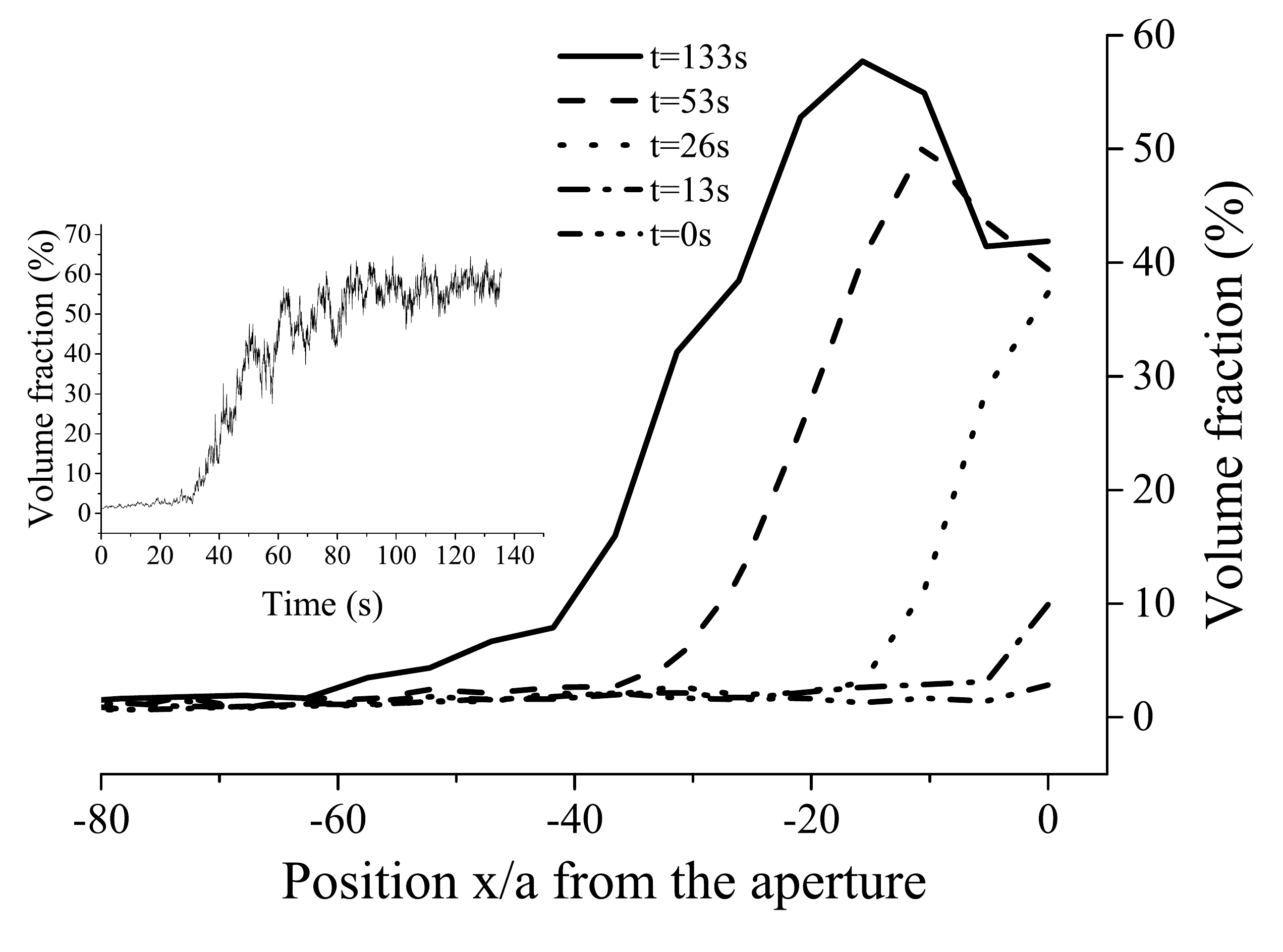}
 \end{center}
\caption{Volume fraction of cells as a function of the upstream distance $x/a$ from the constriction. The aperture is set as the $x$ origin. Measurements of volume fraction are done indirectly by measuring the absorbance of the transmitted light averaged over a surface of $\unit{3.7\times 10^3}{\micro\meter\squared}$, equivalent to 140 cells. Inset: Evolution of volume fraction of the cells against time increasing up to a stationary plateau measured at a position $x/a=10$ from the aperture.}\label{absorbance_time}
\end{figure}

In this work, we focus our study on clogging events occurring at a constriction due to cell build-up of microalgae. Understanding behavior of highly concentrated motile micro-organisms and especially their short-range interactions through adhesion or friction remains a challenge. 

In our experiments, we use a negative  phototactic strain of green algae {\it Chlamydomonas reinhardtii} (CR) swimming away from light and moving toward a constriction (fig. \ref{snapshots}). We identify clogging events and analyze the dynamics of evacuation. We show that what looks like a FiS effect might occur in our system supposedly governed by hydrodynamics. 

\subsection{Experimental details}

CR is a biflagellate photosynthetic cell \cite{harris2009chlamydomonas}. Cells are grown under a 14h/10h light/dark cycle at $22\degree$C and are harvested in the middle of the exponential growth phase. Its front flagella beat in a breakstroke manner and propel the cell in the fluid. The swimming is characterized by a persistent random walk in absence of bias \cite{polin2009chlamydomonas,garcia2011random}. However, in the presence of a light stimulus (green wavelength, i.e., around 510 nm), the strain used in this study (CC-124) tends to swim away from the light source and perform a ballistic motion \cite{garcia2013light}, this is known as negative phototaxis. CR are used with no further preparation. The cells are finally introduced within a microfludic device composed of two reservoir chambers of height $\unit{20}{\micro\meter}$  (twice the cell diameter) linked by a bottleneck constriction of width $d=\unit{15}{\micro\meter}$. Microfabrication is made of transparent PDMS by means of soft lithography processes \cite{qin2010soft}. 

We observe the cells under a bright field inverted microscope (Olympus IX71, magnification x2) coupled to a 14 bit CCD camera (Prosilica GX) used at a framerate of 15 fps. The sample is enclosed in an occulting box with two red filtered windows for visualisation in order to prevent the microscope light from triggering phototaxis.

The algae are initially distributed homogeneously in the chamber. A white LED light is switched on on the left side of the sample. Negative phototaxis make the microswimmers swim towards walls and at the constriction. Particle tracking is performed using the library Trackpy \cite{trackpy, crocker}. The volume fraction of cells when very dense regimes are reached is deduced from grey level intensity measurements that were calibrated beforehand.

Figure \ref{snapshots} illustrates one typical geometry of the experiments. The top sequence shows a series of microscopy images at different timescales following the light stimulus. The microswimmers swim away from the light and toward the constriction (from left to right). The colormaps below represent the volume fraction distribution deduced from grey level intensity measurements evaluated over a box sizes of about 10 micrometers and averaged over 3 seconds. The density is observed to increase near the constriction and to eventually build up an arche-like pattern embracing a depletion zone at the constriction.

\subsection{Results}
\begin{figure}[!ht]
  %\centering
  \begin{center}

    \begin{tabular}{l}
    a.%hspace*{-0.8cm}
\includegraphics[width=\columnwidth]{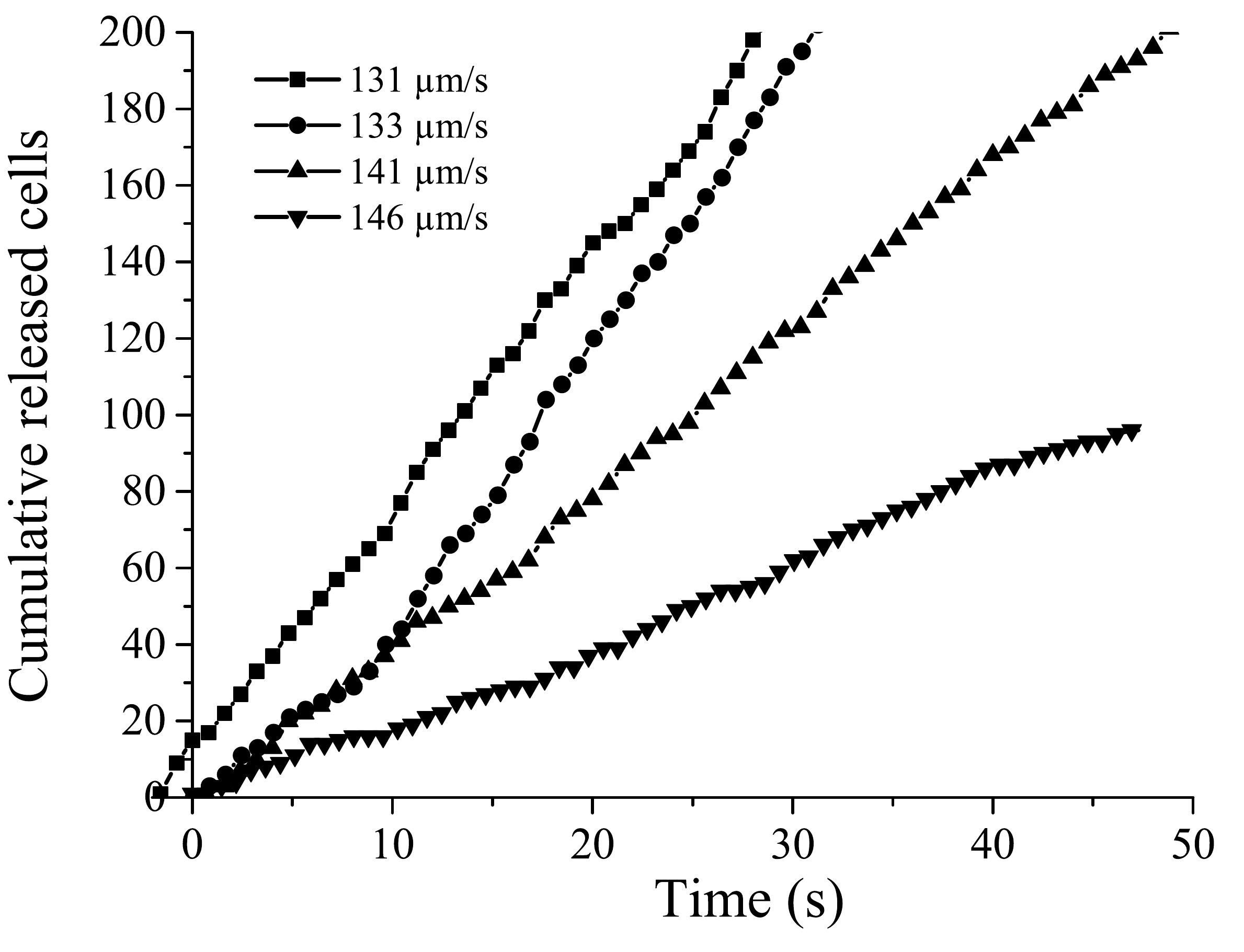}\\  
b.%hspace*{-0.8cm}
\includegraphics[width=\columnwidth]{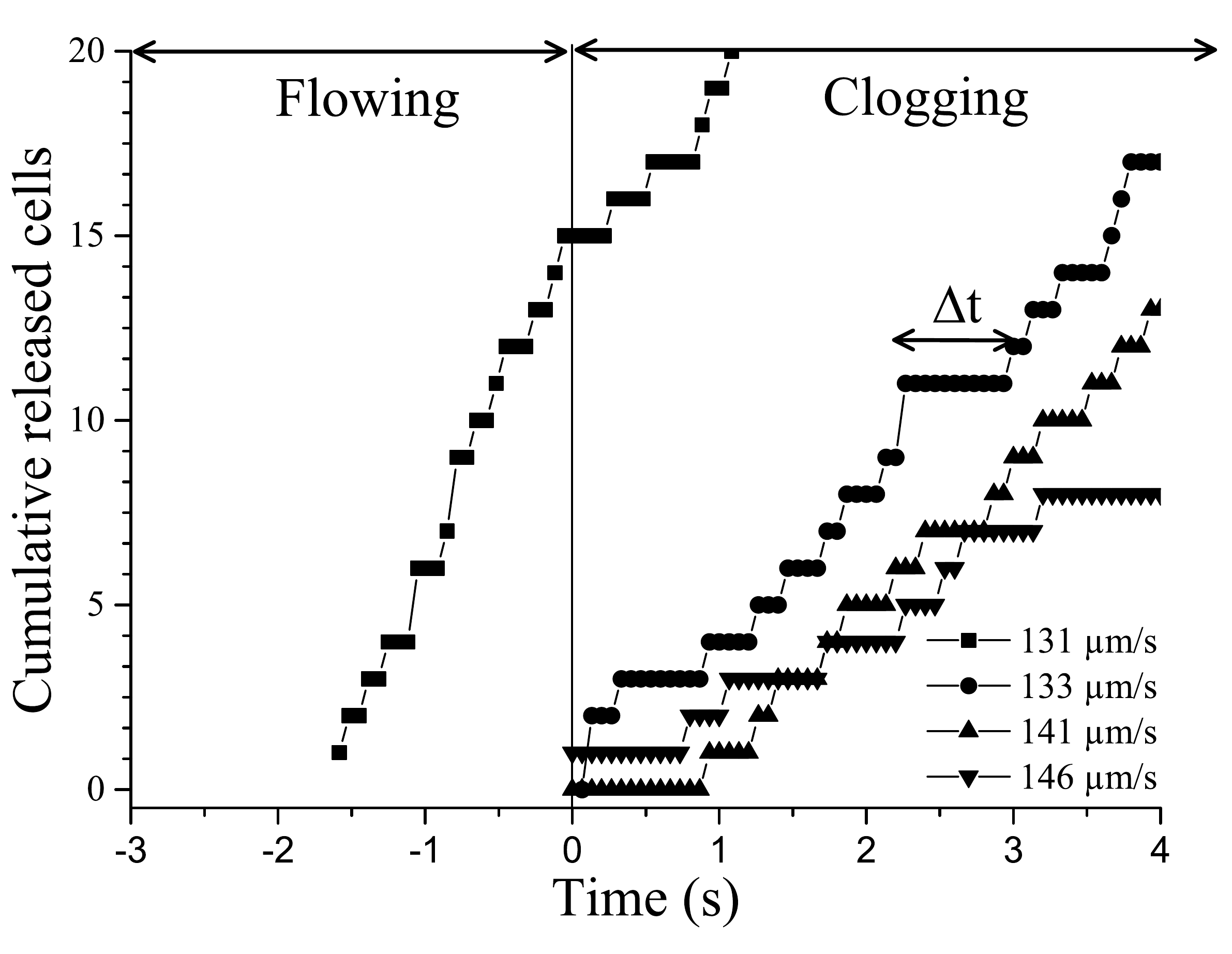}
    \end{tabular}
 \end{center}
\caption{a. Cumulative number of cells released at the exit for different swimming velocities against time. b. Zoom of the previous graph. $\Delta t$ defines the time lapse between two effective releases. Clogging are effective when clog duration is more than $\tau=v/a$,   Here, the time 0 has been set to be the starting time for the observation of clogging and clogging events occur for $t>0$ s.}\label{comptage_cumule}
\end{figure}

The density of cells that is obtained after few tens of seconds is very high reaching up to values close to $60\%$. The dynamics of evacuation is characterized by intermittent bursts of cells separated by blocages where no algae escape from the constriction. Initially, cells are forming two waiting lines on either side of the constriction ($x=0$) toward which they are heading. During this first phase that we call "flowing regime", no clogging event is observed and cells are escaping one by one  the bottleneck constriction from the two lines separated by a time interval of the order of $\tau$. We can estimate $\tau$ as $\tau=a/V_0$, where $a \approx 10\mu m$ is the size of a cell while $V_0 \approx \unit{130}{\micro\meter\per\second}$ is the swimming mean velocity of algae. When no clogging occurs, $\tau \approx \unit{100}{\milli\second}$. Then, algae accumulate in front of the constriction and the cell volume fraction starts to increase as shown on Figure \ref{absorbance_time}. Algae build up during approximatively $80\, sec.$, then the volume fraction (averaged over an area of $\unit{3.7\times 10^3}{\micro\meter\squared}$ just before the aperture) saturates, reaching $55\%$. The volume fraction is measured at each relative position $x$ from the aperture with a $\unit{52}{\micro\meter}$ increment, averaged over an area of $\unit{3.7\times 10^3}{\micro\meter\squared}$. This volume fraction profile is plotted at different times in Figure \ref{absorbance_time}. The profile is uniform at the beginning of the build-up process ($t=0$ s). Then, it first increases monotonously with $x$, with a maximum density of cells at the constriction. After $50 s$, the volume fraction in cells appears to be non-monotonous with $x$: a higher density front moves from the constriction ($x=0$) to roughly at $200 \mu m$  upstream the constriction  where it reaches $60\%$. Remarkably, this is also observed in experiments \cite{garcimartin2017pedestrian} and numerical simulations of pedestrians \cite{parisi2015faster}, the maximum of density appears roughly around ten agent sizes before the exit, thus revealing the formation of an arch structure at the origin of the clogging. In the case of CR, the maximum density upstream appears at around 20 diameter of cells.

When cells are  blocked by their neighbors at the constriction, a blocage occurs during a time $\Delta t > \tau$. Here, we study the statistical distribution of times lapses $\Delta t$ of blocages. In figure \ref{comptage_cumule}, we plotted the cumulative number of cells going through the constriction over time $N(t)$. This curve yields insights on the intermittency flux of particles at the exit for different swimming velocities. We note that time lapses of blocage ($\Delta t$) appear in the clogging regime and last several $\tau$, thus creating intermittent cells bursts through the bottleneck.

This clogging situation can be quantified by computing the complementary cumulative distribution of exit times $\Delta t$, the so-called survival function. It represents the probability to find a time duration of a blocage larger than $\Delta t$. This function is used to eliminate or reduce the noise in the probability distribution function $\rho(\Delta t)$. The survival function is thus defined as :
\begin{align*}
P(t>\Delta t)=\int_{\Delta t}^\infty \rho(\Delta t') d\Delta t'.
\end{align*}
The survival function is shown to follow a power law decay with an exponent equal to $1-\alpha$ where $\alpha$ is such as $\rho(\Delta t) \sim \Delta t^{-\alpha}$  similar to what is reported in the literature concerning the dynamics of several systems, such as humans, mice, sheep, granular media and colloidal suspensions \citep{zuriguel2014clogging}. Usually, when no clogging occurs, an exponential decay is observed with a characteristic time $\tau$ since each cell crossing the bottleneck at an average frequency $\sim \tau^{-1}$ can be treated statistically as a queueing problem with independent events. Here, the power law decay of the survival function illustrates the clogging nature of the congestion of cells. The value of the exponent $\alpha$ can be viewed as a kind of efficiency of evacuation: large values of $\alpha$ may indicate faster evacuation.

\begin{figure}[!ht]
\centering
 % \hspace*{-1cm}
\includegraphics[width=\columnwidth]{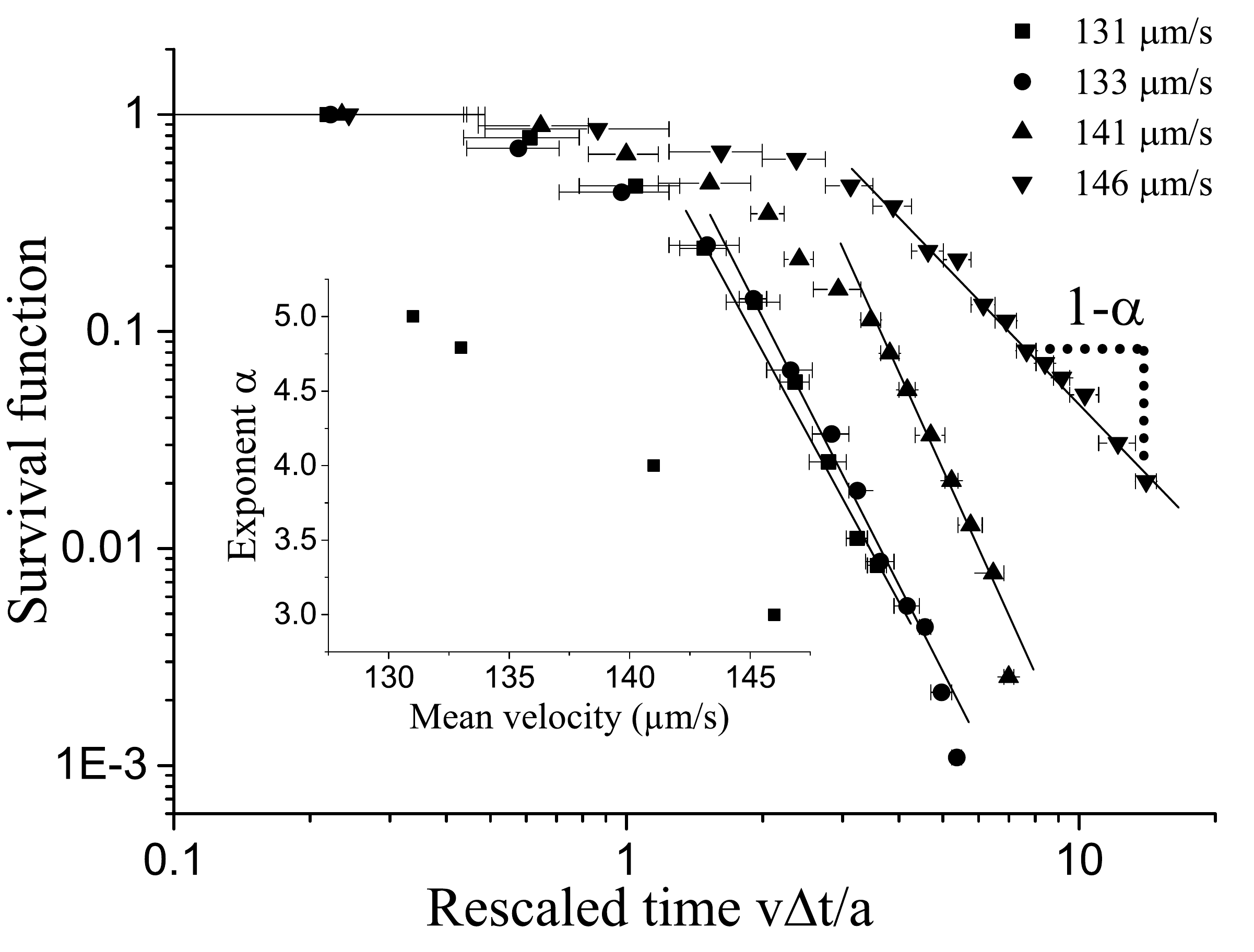}  	   
  \caption{Cumulative probability distribution of time lapses (survival function) for several mean swimming velocities: they all decrease in a power-law fashion. Time is normalized by $\tau=a/v$. Inset: slope of the power law distribution as a function of the mean swimming velocity.}  \label{survie_exp}
\end{figure}

In Fig. \ref{survie_exp}, we plotted survival fonctions for different mean velocities of swimming corresponding to different experiments. The velocity is measured far upstream the constriction and its mean value is found to vary slightly from one culture to an other one. We probe the effect of this small variability (about 12$\%$  in the swimming velocity) on the decay of the nature of the clogging. Figure \ref{survie_exp} shows that as the evacuation velocity increases, the probability to get larger time during which flux is interrupted increases. The  exponent $\alpha$ is plotted as a function of the velocity in the inset of fig.\ref{survie_exp} and is found in an interval of values such as $3<\alpha<5$. Remarkably, the exponent $\alpha$ is found to be very sensitive to modest variations of velocities.

\subsection{Discussion}

%This experiment highlights the importance of contacts and friction between cells since without it, no FiS effect can occur \cite{helbing2000simulating, parisi2007faster}. Friction governs the obstruction of flow through a narrow passage in many systems irrespective of their scale such as granular media, crowd of panicking people or herd of animals \cite{garcimartin2017pedestrian,garcimartin2015flow,zuriguel2005jamming}. In our case, if the real nature of contacts between neighboring CR is not well understood, it seems that short range adhesion between cells is possible through their flagella \cite{kreis2018adhesion}. 

%These  different crowd systems in such an evacuation process show the same evolution: a fast increase of the density while approaching the aperture, then a slight decrease after a maximum.

%
%We also measure the burst size distribution, {\it i.e.} the number of cells per burst. This distribution (Fig. \ref{burst_size_distribution}) looks close to an exponential decay as it was previously observed for crowds, illustrating once again the remarkable similarity of the two phenomena.  The distribution decays faster when cell velocity increases, illustrating that the size of bursts are shorter on average when cells are faster.
%
%
%\begin{figure}[!ht]
%\centering
%  \hspace*{-1.5cm}
%  \includegraphics[width=10 cm]{./figures_pdf/burst_size_distribution}  	   
%  \caption{Probability distribution of burst size (number of cells per burst) for several mean swimming velocities.}  \label{burst_size_distribution}
%\end{figure}

In this work, we analyze, in the framework of crowd dynamics phenomenology, a congestion at a bottleneck of motile cells  ({\it Chlamydomonas reinhardtii}) swimming away from a light source. This experiments allowed us to reach new regimes of very dense active particles that show interesting features such as clogging events, arches-like pattern of density and paradoxical evacuation dynamics that may ressemble Faster is Slower effect.

A FiS effect represents the increase of the probability of finding larger time laps $\Delta t$ when particles are faster \cite{zuriguel2014clogging}. This effect has been suspected to find its origin in formation of arches favored by tangential frictions \cite{helbing2000simulating}. These arches may be more stable when loaded like architectural arches, and might be shattered by loads exerted in a different direction than those driving the flow. In that respect, the jammed CR here might be viewed as an example of fragile active matter after a concept originally introduced for granular jammed material  \cite{cates1998jamming} .  Nevertheless,  it has been also shown that FiS can also be obtained in models deprived of friction between particles \cite{faure2015crowd}. This raises the question of whether a FiS effect can persist in the presence of hydrodynamic forces or whether its manifestation indicates that we are facing frictional forces in the particular case of jammed active matter. Indeed contact forces have been shown to play a dominant role in the rheology of dense passive suspensions \cite{seto2013discontinuous,mari2014shear}.

A route for a better investigation of FiS effect in microswimmers crowds is to analyze the effect of an upstream obstacle. Although debated in the literature \cite{Garcimart_n_2018}, it has been suggested \cite{helbing2000simulating,  zuriguel2014clogging} that if the system presents a Faster is Slower effect, the presence of an upstream obstacle may slow down on average the flow of agents through the bottleneck, which consequently decreases the time of an evacuation. The role of an obstacle is also to regularize the flow around it and therefore can facilitate the evacuation and reduces very much the clogging effect. This will be quantified together with the effect of the size of the constriction. Indeed, here the width of the constriction has been designed to be small enough in order to observe clogging (around one and a half microswimmer's diameter) and has been found to be smaller than what is usually observed in standard crowds.

For a congestion of microswimmers, the measure of $\alpha$ exponent could be a way to characterize the friction between cells at high densities. We believe that our work could pave the way to future investigations to better understand cell-cell close interactions. Indeed, such experiments can shed light on the debate about the minimum ingredients needed to observe what resembles universal behaviors. In particular, this will need to  analyze the importance of the nature of frictions and contacts between motile cells in the evacuation dynamics.

\begin{acknowledgments}
We thank Alexandre Nicolas for insightful discussions. We thank the French-German university program "Living Fluids" (grant CFDA-Q1-14) (M. B.-C.-B., P.P. and S.R.) and the Programme 80Prime of CNRS. \end{acknowledgments}

%%%%%%%%%%%%%%%%%%%%%%%%%%%%%%%%%%%%%%%%%%%%%%%%%%%%%%%
%%%%%%%%%%%%%%%%%%%%%%%%%%%%%%%%%%%%%%%%%%%%%%%%%%%%%%%
%%%%%%%%%%%%%%%%%%%%%%%%%%%%%%%%%%%%%%%%%%%%%%%%%%%%%%%

\revappendix*

\end{document}

%% file: author_list.tex
\author{Marvin Brun-Cosme-Bruny}
%\thanks{}
\affiliation{Univ. Grenoble Alpes, CNRS, LIPhy, F-38000 Grenoble, France}

\author{Vincent Borne}
%\thanks{}
\affiliation{Univ. Grenoble Alpes, CNRS, LIPhy, F-38000 Grenoble, France}
%\thanks{}

\author{Sylvain Faure}
%\thanks{}
\affiliation{Univ. Grenoble Alpes, CNRS, LIPhy, F-38000 Grenoble, France}

\author{Bertrand Maury}
%\thanks{}
\affiliation{Univ. Grenoble Alpes, CNRS, LIPhy, F-38000 Grenoble, France}

%\altaffiliation{Corresponding author}
\author{Philippe Peyla}
\email[]{philippe.peyla@univ-grenoble-alpes.fr}
%\thanks{}
\affiliation{Univ. Grenoble Alpes, CNRS, LIPhy, F-38000 Grenoble, France}

\author{Salima Rafa\"{\i}}
\email[]{salima.rafai@univ-grenoble-alpes.fr}
%\thanks{}
\affiliation{Univ. Grenoble Alpes, CNRS, LIPhy, F-38000 Grenoble, France}